\documentclass[twocolumn,showpacs,prl,aps]{revtex4}
\usepackage{amsmath}
\usepackage{amsfonts}
\usepackage{amssymb}
\usepackage{graphicx}
\usepackage{hyperref}
\usepackage{color}
\usepackage{bm}

\everymath{\displaystyle} 
\newcommand{\pr}[1]{\ensuremath{\left[#1\right]}} 
\newcommand{\pc}[1]{\ensuremath{\left(#1\right)}} 
\newcommand{\px}[1]{\ensuremath{\left\lbrace#1\right\rbrace}} 
\newcommand{\bra}[1]{\ensuremath{\left\langle#1\right\vert}} 
\newcommand{\ket}[1]{\ensuremath{\left\vert#1\right\rangle}} 
\newcommand{\md}[1]{\ensuremath{\left\vert#1\right\vert}} 
\begin{document}

\title{Casimir-Polder interaction from exact diagonalization and
state mixing near surfaces}
\author{Sofia Ribeiro$^{1}$, Stefan Yoshi Buhmann$^{2,3}$, Thomas
Stielow$^4$
and Stefan Scheel$^{4}$}
\affiliation{$^{1}$ School of Physics and Astronomy, University of
Southampton, Southampton, SO17 1BJ, United Kingdom}
\affiliation{$^{2}$ Physikalisches Institut,
Albert-Ludwigs-Universit\"at Freiburg, Hermann-Herder-Str. 3, 79104
Freiburg, Germany}
\affiliation{$^{3}$ Freiburg Institute for Advanced Studies,
Albert-Ludwigs-Universit\"at Freiburg, Albertstr. 19, 79104 Freiburg,
Germany}
\affiliation{$^{4}$ Institut f\"ur Physik, Universit\"at Rostock,
Universit\"atsplatz
3, D-18051 Rostock, Germany}

\begin{abstract}
Dispersion forces have a sizeable effect on the energy
levels of highly excited Rydberg atoms when brought close to material
surfaces. Rydberg atoms experience energy shifts in the GHz range
at micrometer distances, suggestive of considerable state admixture.
We show that despite the non-applicability of perturbation
theory for Rydberg atoms near a surface, the energy shift due to the
dispersion interaction can be obtained from an exact diagonalization
of the interaction Hamiltonian by finding the zeros of the Pick
function. Moreover, we show that contrary to intuition from single-mode 
approaches, surface-induced state mixing is generally suppressed even
for large interaction energies. We give a tailored example where
mixing is observable despite this effect.
\end{abstract}

\pacs{12.20.-m, 42.50.Nn, 32.80.Ee}

\maketitle

\section{Introduction}

The change of atomic properties due to interactions with the
quantized electromagnetic field in the presence of macroscopic bodies
is a well-known effect. The theoretical approaches to the body-induced
atomic (Casimir-Polder) energy shift are usually based on
second-order perturbation theory \cite{Wylie85,acta2008}. A perturbative 
framework tacitly assumes the effect to be small, i.e., that the
energy levels and their differences are not changed significantly:
\begin{equation}
\md{\bra{m}\hat{H}_{\mathrm{int}} \ket{n}} \ll \md{E^0_m -E^0_n}.
\end{equation}
Valid for ground-state and lowly-excited atoms, this assumption breaks
down for Rydberg atoms as the dipole-matrix elements can become
huge and the spectrum of neighbouring excited states is very dense.

Research in the field of Rydberg atoms has recently seen an enormous
resurgence due to their extraordinary properties and the technological
advances in their coherent manipulation. This has resulted in
proposals and, in part, implementation of photonic quantum devices
such as quantum gates and single-photon sources
\cite{PRL102_240502_2009,PRL_87_3_2001,PRA66_065403_2002,%
Science336_887_2012,PRA87_053412_2013}, quantum computers
\cite{PRL104_010503_2010,PRL104_010502_2010} and quantum simulators
\cite{NaturePhys6_382_2010}, ideas that could also be expanded to
Rydberg states of ions \cite{NJP13_075014_2011}. In all these
situations, precision control of atomic systems requires
well-understood surface effects.

The term Rydberg atom refers to an atom with one or more of its
valence electrons in a state with large principal quantum number. Such
states are relatively long-lived \cite{Gallagher}. Their large radius
gives Rydberg states gigantic dipole transition moments, resulting in
strong long-range dipole-dipole interactions. The subsequent van der
Waals dispersion interaction prevents the multiple excitation of
Rydberg states in atomic ensembles, an effect known as dipole (or
Rydberg) blockade \cite{NPL5_110_2009,NPL5_115_2009}.
 
Similarly, due to their large electric polarizability the
Casimir-Polder force on atoms in Rydberg states near a surface is
several orders of magnitude larger than the corresponding force on
atoms in their ground state. This force scales with the principal
quantum number $n$ as $n^4$ \cite{AlexPRA2010}.
For this reason, experimental investigation into atom-surface
interactions have often used Rydberg atoms
\cite{PRA37_9_1988,Sandoghdar92,Brune94,Sandoghdar96,Marrocco98,Kuebler10}. 
In the example of Ref.~\cite{PRA37_9_1988}, highly excited Cs and Na atoms were 
passed through a cavity of gold-coated mirrors and the transmission
was measured. Due to the attractive atom-surface interaction, a
deflection towards the surface was observed. The interaction strength
for atoms prepared in states with principal quantum numbers $n = 12 -
30$ was shown to be $3 - 4$ orders of magnitude larger than those for
ground-state atoms.

Such extreme body-induced level shifts naturally fall into a regime
where perturbation theory breaks down. This is shown in
Fig.~\ref{Fig:ExactRes} where we compare the Casimir-Polder shift
(black dashed curves) with the atomic unperturbed energies (red
lines). Instead, one has to find the eigenvalues of the matrix
interaction
\begin{equation}
W = \bra{\psi}  \hat{H} \ket{\psi}
\end{equation}
in a suitably chosen basis. 
In the past, Bogoliubov-type transformations have been used to exactly 
diagonalize the atom-field Hamiltonian modelling the atoms as three 
dimensional isotropic harmonic oscillators interacting with the quantum 
electromagnetic field. In this way, both the atom-atom interaction
\cite{PRA72_052106_2005} and atom-surface interaction have been
derived \cite{PRA85_062109_2012}.

Strong coupling of a single cavity mode $\nu$ with an excited atom
can be described in relatively simple terms on the basis of the
Jaynes-Cummings model. Here, an exact diagonalization is readily
available \cite{Jaynes63} where the eigenstates of the strongly
coupled system are superpositions of the type
$\alpha|1\rangle|0_\nu\rangle+\beta|0\rangle|1_\nu\rangle$
\cite{Haroche91,Englert91}. For sufficiently large interaction
energies, an initially excited atom will hence acquire a considerable
mixing amplitude $\beta$ when interacting with the cavity mode. As
we will show, the situation is drastically different for
the broad-band coupling of an atom with a single surface.

In this article, we develop a non-perturbative description of the strong
interaction of a multilevel Rydberg atom with a nearby surface.
This is achieved by exploiting the arrowhead form of the interaction
Hamiltonian to find its exact eigenvalues as zeros of the Pick
function \cite{arrowhead_paper}. We obtain results for both energy
shifts and mixing amplitudes, showing that the latter typically
deviate from the above intuition from resonant single-mode coupling.

\section{Macroscopic quantum
electrodynamics}\label{sec:degeneratePTbasic}

In electric dipole approximation, the Hamiltonian that governs the dynamics of 
the coupled atom-field system can be written as \cite{acta2008}
\begin{gather}
\hat{H}
= \hat{H}_{F} + \hat{H}_{A} + \hat{H}_{\mathrm{int}} \nonumber \\
= \int\limits_0^\infty d \omega \int  d^3r \, \hbar \omega\, 
\hat{\mathbf{f}}^\dagger(\mathbf{r}, \omega)\cdot
\hat{\mathbf{f}}(\mathbf{r},\omega) + \sum_n \hbar \omega_n \hat{A}_{nn}
\nonumber \\
- \sum_{m,n} \hat{A}_{nm} \mathbf{d}_{nm} \cdot \hat{\mathbf{E}}(\mathbf{r}_A)
\,.\label{eq:hamiltonian}
\end{gather}
Here, $\hat{H}_{F}$ is the Hamiltonian of the medium-assisted electromagnetic 
field which is expressed in terms of a set of bosonic variables 
$\hat{\mathbf{f}}^\dagger(\mathbf{r},\omega)$ and 
$\hat{\mathbf{f}}(\mathbf{r},\omega)$ that are interpreted as amplitude 
operators for the elementary excitations of the medium-field system 
(polaritons). The Hamiltonian $\hat{H}_{A}$ of the uncoupled atom can be 
expanded in terms of its eigenenergies $E_{n}=\hbar\omega_n$ and eigenstates 
$\ket{n}$, where $\hat{A}_{nm}= \ket{n} \bra{m}$ denotes the transition 
operators between two internal atomic states. The atom-field interaction is 
given in terms of dipole transition matrix elements 
$\mathbf{d}_{nm}=\bra{n}\hat{\mathbf{d}}\ket{m}$ and the electric-field operator
\begin{gather}
\hat{\mathbf{E}}(\mathbf{r}_A) =  \int_0^\infty   d \omega  \int d^3r\, 
\left[ \mathbf{G}_e(\mathbf{r}_{A}, \mathbf{r},\omega) \cdot 
\hat{\mathbf{f}}(\mathbf{r},\omega) + \mbox{h.c.} \right].
\end{gather}
The tensor $\mathbf{G}_e(\mathbf{r}_{A}, \mathbf{r}, \omega)$ is related to the 
classical Green tensor $\mathbf{G}(\mathbf{r}_{A},\mathbf{r},\omega)$, i.e. the 
solution of the Helmholtz equation with the appropriate boundary conditions, by
\begin{equation}
\mathbf{G}_{e}(\mathbf{r}, \mathbf{r}', \omega) = i \frac{\omega^{2}}{c^{2}} 
\sqrt{\frac{\hbar}{\varepsilon_{0} \pi} \mathrm{Im}\, 
\varepsilon(\mathbf{r}',\omega)} \, \mathbf{G}(\mathbf{r},\mathbf{r}',\omega),
\end{equation}
where $\varepsilon(\mathbf{r},\omega)$ is the (complex) dielectric permittivity 
of the macroscopic body.

\section{Exact diagonalization: ground-state two-level atom}

Let us consider first a two-level atom with energy levels $\ket{1}$ and 
$\ket{0}$, for which the dipole operator is
$\hat{\mathbf{d}} = \mathbf{d} \hat{A}_{10} + \mathbf{d}^\ast \hat{A}_{01}$. 
The atomic Hamiltonian is thus
\begin{equation}
\hat{H}_{A} = \hbar \omega_{0} \hat{A}_{00}  + \hbar \omega_{1} \hat{A}_{11}.
\end{equation}
It is useful to introduce position-dependent photon-like amplitude operators 
$\hat{a} \pc{\mathbf{r}, \omega}$ and $\hat{a}^{\dagger}\pc{\mathbf{r}, \omega}$ 
as \cite{PRA77_012110_2008}
\begin{eqnarray}
\hat{a} \pc{\mathbf{r}, \omega} = - \frac{1}{\hbar g \pc{\mathbf{r}, \omega} } 
\int d^3 r' \mathbf{d} \cdot \mathbf{G}_{e} \pc{\mathbf{r}, \mathbf{r}', \omega}
\cdot \hat{\mathbf{f}} \pc{\mathbf{r'}, \omega}
\end{eqnarray}
with a normalization factor
\begin{equation}
\label{eq:normalization}
g \pc{\mathbf{r}, \omega} = \sqrt{\frac{\mu_{0}  \omega^2}{\hbar \pi} 
\mathbf{d} \cdot \mathrm{Im} \, \mathbf{G}\pc{\mathbf{r}, \mathbf{r}, \omega}
\cdot \mathbf{d}^\ast}\,.
\end{equation}
The creation operator $\hat{a}^{\dagger} \pc{\mathbf{r}, \omega}$ can be used 
to define single-quantum excitations from the ground state $\ket{\px{0}}$ of the 
medium-assisted electromagnetic field, $\ket{\mathbf{r}, \omega} = 
\hat{a}^{\dagger} \pc{\mathbf{r}, \omega} \ket{\px{0}}$. These ladder operators 
obey the usual commutation rule $\pr{\hat{a} \pc{\mathbf{r}, 
\omega},\hat{a}^{\dagger} \pc{\mathbf{r}, \omega'}} = \delta \pc{\omega - 
\omega'}$. The states $\ket{\mathbf{r}, \omega}$ are eigenstates of the field 
Hamiltonian $\hat{H}_{F}$ such that
\begin{gather}
\hat{H}_{F} \ket{\mathbf{r}, \omega} = \hbar \omega \ket{\mathbf{r}, \omega}.
\end{gather}
The interaction Hamiltonian can then be re-written as 
\begin{gather}
\hat{H}_{\mathrm{int}} = \hbar \int\limits_{0}^{\infty} d \omega \pr{ g
\pc{\mathbf{r}_{A}, \omega} \hat{a} \pc{\mathbf{r}_{A}, \omega} + g^\ast
\pc{\mathbf{r}_{A}, \omega} \hat{a}^{\dagger} \pc{\mathbf{r}_{A}, \omega} }
\nonumber \\ \times
 \pc{\hat{A}_{01} + \hat{A}_{10}},
\end{gather}
without applying the rotating-wave approximation.

The initial state is taken to be the ground state $\ket{0_A}$ of the two-level 
atom and the vacuum state of the medium-assisted field $\ket{\px{0}}$ which is 
connected via the interaction Hamiltonian to a (continuous) set of final states 
$\ket{1_A}\ket{\mathbf{r}_A,\omega}$ \cite{PR124_6_1961}. Our aim is to 
diagonalize the total Hamiltonian within this basis, that is, to find the exact 
solutions of $\hat{H}|\psi\rangle=\hbar\Omega|\psi\rangle$, where
\begin{equation}
|\psi\rangle = C_{0} \ket{0_A} \ket{\px{0}} + \int_{0}^{\infty} d \omega \, 
C_{1} \pc{\omega} \ket{1_A} \ket{\mathbf{r}_A, \omega}.
\end{equation}
Note that this amounts to truncating the Hilbert space to the zero-
and single-photon sectors. Applying the Hamiltonian to the state
$|\psi\rangle$ yields a set of equations
\begin{subequations}
\begin{align}
\pc{\omega_{0} - \Omega} C_{0}  + \int_{0}^{\infty} d\omega\, C_{1}
(\omega) g(\mathbf{r}_A, \omega) = 0, \label{eq:coef1} \\
g^\ast(\mathbf{r}_A, \omega) C_{0}  + \pc{ \omega +  \omega_1 - \Omega} C_{1}
(\omega) = 0 
\label{eq:coef2}
\end{align}
\label{eq:coefs}
\end{subequations}
for the eigenfrequencies $\Omega$.

For further investigation, we discretize the frequency integral in 
Eq.~(\ref{eq:coef1}) according to
\begin{equation}
\label{eq:discrete}
\int_{0}^{\infty} d\omega\, C_1(\omega) g(\mathbf{r}_A, \omega) \mapsto 
\sum\limits_{i=1}^N \Delta\omega\,C_1(\omega^{(i)}) 
g(\mathbf{r}_A,\omega^{(i)}).
\end{equation}
This leads to a set of $N+1$ equations
\begin{subequations}
\begin{align}
\pc{\omega_{0} - \Omega} C_{0}  + \Delta\omega \sum_{i=1}^N C_{1}(\omega^{(i)}) 
g(\mathbf{r}_A, \omega^{(i)}) = 0,  \\
g^\ast(\mathbf{r}_A, \omega^{(i)}) C_{0}  + (\omega^{(i)}+\omega_1-\Omega) 
C_{1}(\omega^{(i)}) = 0,
\end{align}
\end{subequations}
which can be brought into the matrix form
\begin{equation}
\label{eq:eigenvalues}
\mathcal{M}\cdot\mathbf{c}=0
\end{equation}
with an $(N+1)$-dimensional coefficient vector 
\begin{equation}
\mathbf{c}=(C_0,C_1^{(1)},\ldots,C_1^{(N)})^T. \nonumber
\end{equation}
The matrix  
\begin{equation}
\mathcal{M} = 
\begin{pmatrix}
  a & r_1 & r_2 & \cdots & r_N \\
  c_1 & d_1 & 0 & \cdots & 0 \\
  c_2 & 0 & d_2 & \cdots & 0 \\
  \vdots  & \vdots& \vdots  & \ddots & \vdots  \\
  c_N & 0 & \cdots & \cdots & d_N
 \end{pmatrix}
 \label{eq:matrixM}
\end{equation}
is of arrowhead type with entries
\begin{subequations}
\label{coefficients}
\begin{gather}
a = \omega_0-\Omega \,,\quad 
d_i = \omega^{(i)}+\omega_1-\Omega\,,\\
r_i = \Delta\omega g(\mathbf{r}_A, \omega^{(i)}) \,,\quad
c_i = g^\ast(\mathbf{r}_A, \omega^{(i)}) \,.
\end{gather}
\end{subequations}

The eigenvalue equation \eqref{eq:eigenvalues} has a unique solution if the 
determinant of $\mathcal{M}$ vanishes, where the latter takes the form
\begin{multline}
\det \mathcal{M} = (\omega_0-\Omega) \prod_{i=1}^N 
(\omega^{(i)}+\omega_1-\Omega) \\ -\sum\limits_{i=1}^N \Delta\omega 
|g(\mathbf{r}_A, \omega^{(i)})|^2 \prod_{j\ne i} (\omega^{(i)}+\omega_1-\Omega)= 
0
\end{multline}
and is known as the Pick function \cite{arrowhead_paper}. The solution
to the above equation is
\begin{equation}
(\omega_0-\Omega) - \sum\limits_{i=1}^N \Delta\omega
\frac{|g(\mathbf{r}_A,\omega^{(i)})|^2}{\omega^{(i)}+\omega_1-\Omega}
=0.
\end{equation}
Performing the continuum limit by reversing relation \eqref{eq:discrete} and 
inserting the normalization $g(\mathbf{r}_A, \omega)$, 
Eq.~\eqref{eq:normalization}, we find
\begin{equation}
\label{eq:eigenvalue}
\omega_0 - \Omega = \frac{\mu_0}{\hbar \pi}
\int_{0}^{\infty} d \omega \, \frac{\omega^2\mathbf{d}\cdot\mathrm{Im}\,
\bm{G}(\mathbf{r}_A,\mathbf{r}_A,\omega)\cdot\mathbf{d}^\ast} 
{\omega+\omega_1-\Omega}.
\end{equation}
This is a transcendental equation for the eigenvalue $\Omega$. 

Writing $\Omega=\omega_0+\delta\omega_0$ and noting that 
$\omega_A=\omega_1-\omega_0$ is the (unperturbed) atomic transition frequency, 
Eq.~\eqref{eq:eigenvalue} assumes its more familiar form
\begin{equation}
\label{eq:nonperturbative}
\delta\omega_0 = -\frac{\mu_0}{\hbar \pi}
\int_{0}^{\infty} d \omega \, \frac{\omega^2\mathbf{d}\cdot\mathrm{Im}\, 
\bm{G}(\mathbf{r}_A,\mathbf{r}_A,\omega)\cdot\mathbf{d}^\ast} 
{\omega+\omega_A-\delta\omega_0}.
\end{equation}
Only if the energy shift $\delta\omega_0$ is much smaller compared to the 
transition frequency $\omega_A$, $\delta\omega_0\ll\omega_A$, does 
Eq.~\eqref{eq:nonperturbative} revert to its perturbative form without 
$\delta\omega_0$ in the denominator under the frequency integral
\cite{Wylie85,acta2008}.

The exact diagonalization also yields the normalized eigenvector of the matrix 
\eqref{eq:matrixM} corresponding to the eigenvalue $\Omega$ 
\cite{arrowhead_paper}:
\begin{equation}
\label{v}
v = \frac{x}{ \Vert x \Vert}, \; x = [ 1 ,- r_1/d_1, -r_2/d_2 , \ldots, 
-r_N/d_N ]^T\,.
\end{equation}
Substituting Eqs.~(\ref{coefficients}) for $r_i$ and $d_i$ and performing the 
continuum limit, one finds
\begin{gather}
C_0=\frac{1}{\sqrt{\mathcal{N}}}\,,
\quad
C_1(\omega)=-\frac{1}{\sqrt{\mathcal{N}}}\,
\frac{g(\mathbf{r}_A,\omega)}{\omega+\omega_A-\delta\omega_0}
\end{gather}
with the normalization factor
\begin{gather}
\mathcal{N}=1+ \int_{0}^{\infty}d\omega
\frac{|g(\mathbf{r}_A,\omega)|^2} {(\omega+\omega_A-\delta\omega_0)^2}\,.
\end{gather}
Hence, the probability for the atom to be found in its initial, unperturbed 
ground state when brought close to the surface is
\begin{gather}
p_0 = |C_0|^2 \nonumber \\
 = \Biggl[1 + \frac{\mu_0}{\hbar \pi} \int_{0}^{\infty} 
d \omega \frac{\omega^2\mathbf{d}\cdot\mathrm{Im}\, 
\bm{G}(\mathbf{r}_A,\mathbf{r}_A,\omega)\cdot\mathbf{d}^\ast}
{(\omega+\omega_A-\delta\omega_0)^2}\Biggr]^{-1}.
\end{gather}
On the other hand, there is a finite probability
\begin{equation}
p_1=\int_{0}^{\infty}
d \omega \,|C_1(\omega)|^2
=1-p_0
\end{equation}
for the atom to be transferred to its excited state due to surface-induced 
state-mixing. Alternatively, these probabilities could have been determined 
using the Fano diagonalization method \cite{PR124_6_1961}.

\section{Extension to multilevel Rydberg atoms}

\begin{figure}
 \centering
\includegraphics[scale=.25]{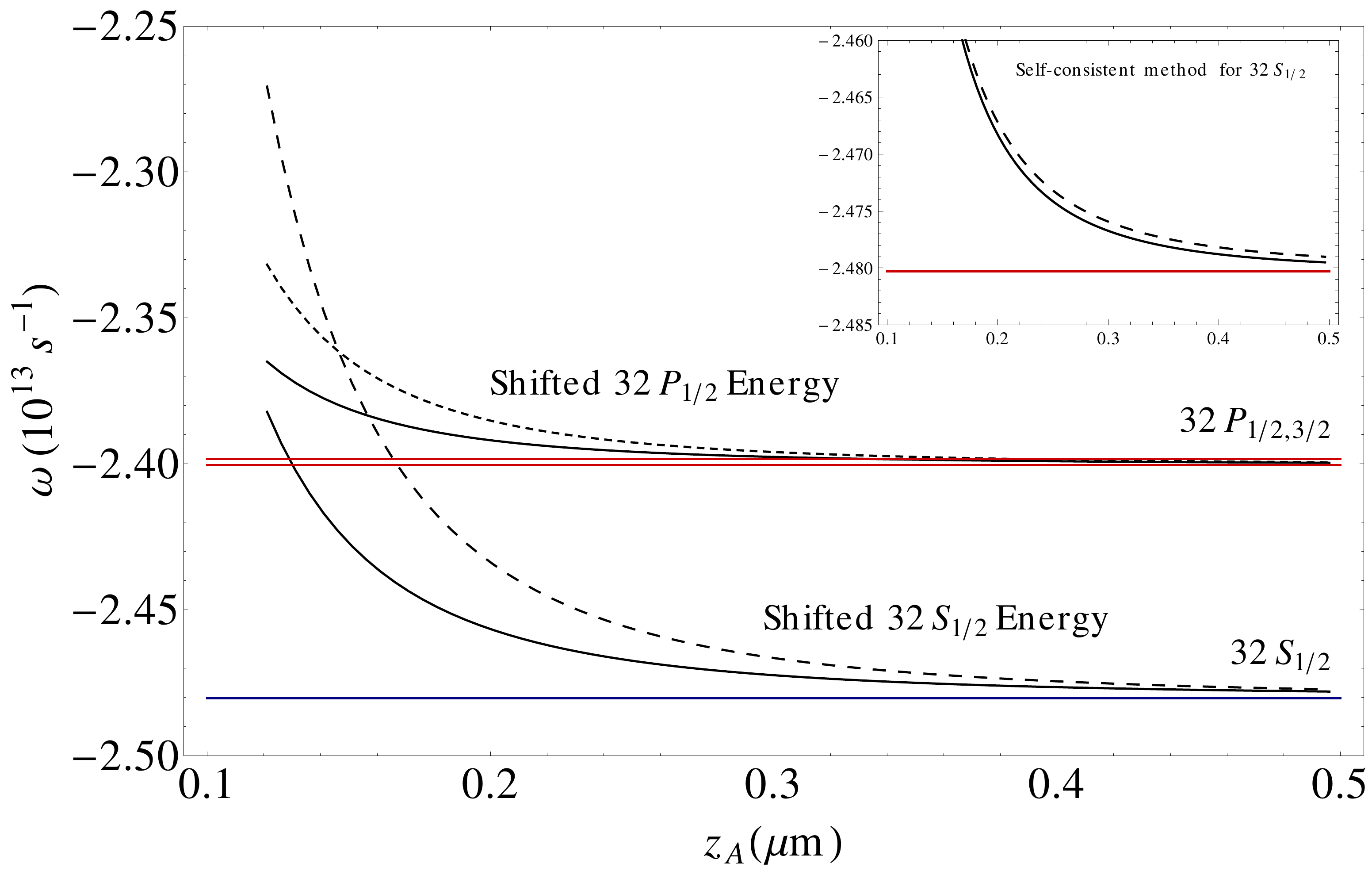}\\
 \caption{(Color online) Change of energy of the 32$S_{1/2}$ and
32$P_{1/2}$ Rydberg states of Rb in the presence of a gold plate
(black solid lines). We compare the self-consistent results with those
from perturbation theory (inset plot in black dashed lines). The substrate
gold was describe by a Drude model $\varepsilon (\omega) = 1 - 
\frac{\omega_{p}^{2}}{\omega (\omega + i \gamma)}$ with
$\omega_{p}=1.37\times 10^{16}\mbox{s}^{-1}$ and $\gamma = 4.12
\times 10^{13} \mbox{s}^{-1}$.
The atomic dipole matrix elements were calculated using wave
functions based on the Numerov method \cite{Blatt81,Bhatti81}. Due to
their sharply peaked nature as a function of $n$, only states with
$|\Delta n|\le 2$ had to be considered.
}  \label{Fig:ExactRes}
\end{figure}

The analysis for a two-level system can easily be extended to multiple
excited states that may be distinguished by suitable quantum numbers. For an 
atom initially prepared in an unperturbed state $\ket{n_A}$, we are seeking 
eigenstates of the form
\begin{multline}
\ket{\psi_n} = C_{n} \ket{n_A}  \ket{\px{0}} \\
+\sum_{k\neq n} \int_{0}^{\infty}   d \omega   \int   d^3r 
\mathbf{C}_{k}(\mathbf{r},\omega)\cdot\ket{k_A} 
\ket{\mathbf{1}(\mathbf{r},\omega)}
\label{psin}
\end{multline}
with $\ket{\mathbf{1}(\mathbf{r},\omega)} 
=\hat{\mathbf{f}}^\dagger(\mathbf{r},\omega)\ket{\px{0}}$. The eigenvalue 
equations then take the form
\begin{subequations}
\begin{align}
&\pc{\omega_{n} - \Omega_{n}} C_{n} \nonumber\\
& - \frac{1}{\hbar}\sum_{k\neq n}  \int_{0}^{\infty}d\omega\int d^3r\,  
\mathbf{d}_{nk}
\cdot  \mathbf{G}_{e} \pc{\mathbf{r}_A,\mathbf{r},\omega}  
\cdot\mathbf{C}_{k}(\mathbf{r},\omega)  = 0, \label{eq:coef1mult} \\
&-\frac{1}{\hbar}\,\mathbf{d}_{kn}\cdot  \mathbf{G}_{e} 
\pc{\mathbf{r}_A,\mathbf{r},\omega}C_{n}  + \pc{\omega +  \omega_k - \Omega_n}  
\mathbf{C}_{k}(\mathbf{r},\omega) = 0.
\label{eq:coef2mult}
\end{align}
\label{eq:coefsmult}
\end{subequations}
We generalize the discretization to the case of the index~$i$ now
running over the set of variables $\{k,\mathbf{r},\omega\}$. The Pick
function then yields an eigenvalue $\Omega_n=\omega_n+\delta\omega_n$
with
\begin{equation}
\label{eq:nonperturbative2}
\delta\omega_n = -\frac{\mu_0}{\hbar \pi}\sum_{k\neq n}
 \mathcal{P} \int_{0}^{\infty}  d \omega
\frac{\omega^2\mathbf{d}_{nk}\cdot\mathrm{Im}
\bm{G}(\mathbf{r}_A,\mathbf{r}_A,\omega)\cdot\mathbf{d}_{kn}}
{\omega+\omega_{kn}-\delta\omega_n}\,.
\end{equation}

In order to interpret the implications of our truncation of the
Hilbert space to the zero- and one-photon sectors, let us compare our
result with that of perturbation theory. If the shift $\delta\omega_n$
is much smaller compared to the transition frequency $\omega_{kn}$,
$\delta\omega_n\ll\omega_{kn}$, one may neglect the term
$-\delta\omega_n$ in the denominator to recover the well-known
leading (second-order) perturbative result \cite{Wylie85,acta2008}.
\begin{equation}
\label{eq:nonperturbative3}
\delta\omega_n^{(1)} = -\frac{\mu_0}{\hbar \pi}\sum_{k\neq n}
 \mathcal{P} \int_{0}^{\infty}  d \omega
\frac{\omega^2\mathbf{d}_{nk}\cdot\mathrm{Im}
\bm{G}(\mathbf{r}_A,\mathbf{r}_A,\omega)\cdot\mathbf{d}_{kn}}
{\omega+\omega_{kn}}\,.
\end{equation}

The next-to-leading order correction can be easily found by applying
a linear Taylor expansion of the right hand side of
Eq.~(\ref{eq:nonperturbative2}) in $\delta\omega_n/\omega_{kn}$:
\begin{multline}
\label{eq:nonperturbative4}
\delta\omega_n^{(2)} = \frac{\mu_0^2}{\hbar \pi^2}\sum_{k\neq n}
 \mathcal{P} \int_{0}^{\infty}  d \omega
\frac{\omega^2\mathbf{d}_{nk}\cdot\mathrm{Im}
\bm{G}(\mathbf{r}_A,\mathbf{r}_A,\omega)\cdot\mathbf{d}_{kn}}
{\omega+\omega_{kn}}\\
\times\sum_{l\neq n}
 \mathcal{P} \int_{0}^{\infty}  d \omega
\frac{\omega^2\mathbf{d}_{nl}\cdot\mathrm{Im}
\bm{G}(\mathbf{r}_A,\mathbf{r}_A,\omega)\cdot\mathbf{d}_{ln}}
{(\omega+\omega_{ln})^2}\,.
\end{multline}
We compare this with the contributions from fourth-order perturbation
theory:
\begin{multline}
\label{fourthorder}
\Delta E=-\sum_k\frac{|\langle n|\hat{H}_\mathrm{int}|k\rangle|^2}
{E_n-E_k}\sum_l\frac{|\langle n|\hat{H}_\mathrm{int}|l\rangle|^2}
{(E_n-E_l)^2}\\
+\sum_{k,l,m}\frac{\langle n|\hat{H}_\mathrm{int}|k\rangle
 \langle k|\hat{H}_\mathrm{int}|l\rangle
 \langle l|\hat{H}_\mathrm{int}|m\rangle
 \langle m|\hat{H}_\mathrm{int}|n\rangle}
 {(E_n-E_k)(E_n-E_l)(E_n-E_m)}\,.
\end{multline}
One observes that the correction~(\ref{eq:nonperturbative4}) as
obtained from expanding the energy shift~(\ref{eq:nonperturbative2})
is equal to the first of the two fourth-order contributions. This
type of contribution factorises. The second fourth-order contribution
does not factorise; it contains two-photon intermediate states. We
conclude that our diagonalisation as obtained from restricting the
field Hilbert space to the zero- and one-photon sectors corresponds
to a complete resummation of factorisable contributions from
perturbation theory to all orders. Note that similar resummations
of certain classes of perturbative contributions are in use in
various fields of physics. For instance, microscopic expansions of
dispersion interactions are based on a resummation of all
factorisable contributions of the Born expansion
\cite{Microscopic1,Microscopic2}, ensuring better convergence
\cite{Golestanian}. Structurally similar resummation methods are
commonly used in quantum field theories in the context of mass
renormalisation \cite{Peskin}. 

Our diagonalisation method is complementary to perturbation theory in
the following sense: when extending our method to the two-photon
sector, one would obtain the non-factorisable fourth-order
contribution of Eq.~(\ref{fourthorder}) above together with all its
factorisable higher-order corrections.

To illustrate our results, Fig.~\ref{Fig:ExactRes} shows the results of the 
exact diagonalization (\ref{eq:nonperturbative}) compared to the results
obtained from second-order perturbation theory. One observes that, for
small distances, perturbation theory does not reproduce the results
from the transcendental result obtained by the exact diagonalization.

Our result~(\ref{eq:nonperturbative2}) resembles that of
Ref.~\cite{PRA70_052117_2004}, obtained by a different
(nonperturbative) method from the dynamical evolution of the
atomic variables. The difference is the absence of a term
$+\delta\omega_k$ from the denominator of
Eq.~(\ref{eq:nonperturbative2}). This is due to the fact that we
evaluate the Pick function separately for each state. To estimate the
error made by neglecting the intermediate-state energy shifts
$\delta\omega_k$ in the denominator, we have iteratively solved the set
of equations (\ref{eq:nonperturbative2}) for $\delta\omega_n$ and its
relevant neighbouring states $\delta\omega_k$. Here, we update the
energy levels at each step of the separation until numerical
convergence is reached. The result is shown by the inset plot in
Fig.~\ref{Fig:ExactRes}. Its difference from our original
result~(\ref{eq:nonperturbative2}) is small compared to that between
both results and standard second-order perturbation theory.

\section{Surface-induced state mixing}

Evaluating the eigenvector~(\ref{v}) for the multilevel case
(\ref{eq:coefsmult}), one finds
\begin{gather}
C_0 = \frac{1}{\sqrt{\mathcal{N}_n}}, \; \mathbf{C}_{k}(\mathbf{r},\omega)
= -\frac{1}{\sqrt{\mathcal{N}_n}}\frac{\mathbf{d}_{nk}\cdot  
\bm{G}_e(\mathbf{r}_A,\mathbf{r},\omega)}
 {\hbar \pc{\omega+\omega_{kn}-\delta\omega_n}}
\end{gather}
with a normalization factor
\begin{multline}
\mathcal{N}_n=1+  \\
\frac{\mu_0}{\hbar \pi}\sum_{k\neq n} \mathcal{P} \int_{0}^{\infty}  d \omega
\frac{\omega^2\mathbf{d}_{nk}\cdot\mathrm{Im}  
\bm{G}(\mathbf{r}_A,\mathbf{r}_A,\omega) \cdot \mathbf{d}_{kn}}
 {(\omega+\omega_{kn}-\delta\omega_n)^2}\,.
\end{multline}
The respective eigenstate~(\ref{psin}) agrees with the perturbative result when 
neglecting the term $\delta\omega_n$ in the denominator --- up to a phase in 
the amplitude $\mathbf{C}_{k}(\mathbf{r},\omega)$. Note that we have 
regularised the frequency integral via a principal value in the generalised form
\begin{equation}
\frac{\mathcal{P}}{(z-a)^2}=\frac{1}{2}\lim_{\epsilon\to 0+} 
\biggl[\frac{1}{(z-a+i\varepsilon)^2}
 +\frac{1}{(z-a-i\varepsilon)^2}\biggr]
\end{equation}
so that $\mathcal{N}_n=1+\sum_{k\neq n}\mathcal{N}_{nk}$ with
\begin{multline}
\label{Nnk}
\mathcal{N}_{nk}=\frac{\mu_0}{\hbar}\,
\theta(\omega_{nk}+\delta\omega_n)\\
\times\bigl[\omega^2\mathbf{d}_{nk}\cdot\mathrm{Re}\,  
\bm{G}(\mathbf{r}_A,\mathbf{r}_A,\omega)\cdot\mathbf{d}_{kn}
 \bigr]'_{\omega=\omega_{nk}+\delta\omega_n}\\
-\frac{\mu_0}{\hbar\pi} \int\limits_{0}^{\infty}  d \xi\xi^2 
\frac{(\omega_{kn}-\delta\omega_n)^2-\xi^2}
 {[(\omega_{kn}-\delta\omega_n)^2+\xi^2]^2}\\
 \times \mathbf{d}_{nk}\cdot  
\bm{G}(\mathbf{r}_A,\mathbf{r}_A,i\xi)\cdot\mathbf{d}_{kn}\,.
\end{multline}

We then find probabilities
\begin{equation}
p_0=|C_0|^2=\frac{1}{\mathcal{N}_n}
\end{equation}
for the atom to remain in its initial state $\ket{n_A}$ and
\begin{equation}
p_k=\mathcal{P}\int_{0}^{\infty}d \omega\int d^3r  
\,|\mathbf{C}_{k}(\mathbf{r},\omega)|^2 =\frac{\mathcal{N}_{nk}}{\mathcal{N}_n}
\end{equation}
for it to be found in a different state due to surface-induced
admixture.

Note that our stationary approach is valid only under adiabatic
conditions similar to those considered in the strong-coupling
scenarios of Refs.~\cite{Haroche91,Englert91}: atom and field are
originally prepared in their uncoupled eigenstates far from the
surface, this separable state being an eigenstate of the total system.
As the atom is slowly brought to the vicinity of the surface, the
coupling is adiabatically increased. The system hence remains in an
eigenstate to end up in the coupled state of our stationary approach. 

Contrary to the intuition from resonant single-mode interactions,
the mixing probabilities vanish for a perfectly conducting plate in
the nonretarded limit: in this case,
$\omega^2\bm{G}(\mathbf{r}_A,\mathbf{r}_A,\omega)$ is independent of
frequency, making both the $\omega$-derivative and the $\xi$-integral
in Eq.~(\ref{Nnk}) vanish. From the mixing probabilities near the
well-conducting gold surface as a function of atom-surface distance
$z_A$, one observes that the presence of the metal surface mixes the
unperturbed eigenstates only weakly, with admixture probabilities on
the order of $10^{-5}$ --- despite the enormous energy shifts shown in
Fig.~\ref{Fig:ExactRes}.

\begin{figure}
 \centering
 \includegraphics[scale=.25]{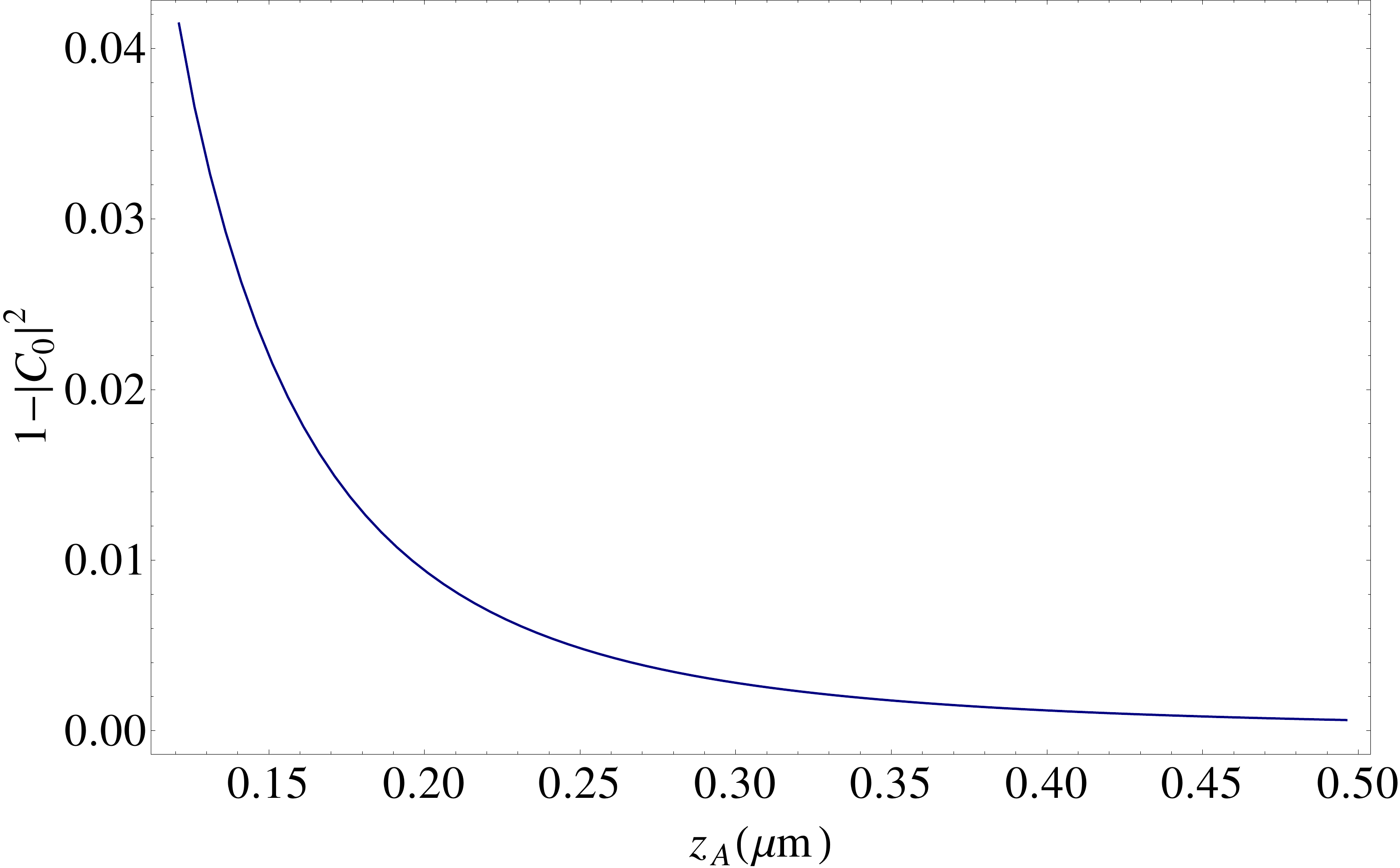}\\
 \caption{(Color online) Probabilities for an $^{87}$Rb atom initially
prepared in the state $32S_{1/2}$ of being in any other excited state
when brought close to a left-handed metamaterial as a function of the
atom-surface distance.}
 \label{Fig:Prob32S}
\end{figure}

Larger admixtures are expected for materials with a strongly
dispersive dielectric response at or below the Rydberg transition
frequencies. As an example, let us discuss the results for a
two-dimensionally isotropic, left-handed metamaterial as described in
Ref.~\cite{APL78_489_2001} via an effective dielectric function,
\begin{equation}
\varepsilon_\mathrm{eff} (\omega) = 1 - \frac{\omega_{p}^2 -
\omega_{0}^2}{\omega^2 - \omega_{0}^2 + i \gamma \omega}
\end{equation}
with $\omega_{p} = 2\pi \times20$~GHz, $\omega_{0} = 2 \pi
\times10$~GHz and $\gamma = 2 \pi \times1$~GHz.
A Rb atom initially prepared in the state $32S_{1/2}$ will be subject
to mixing with states $kP_{1/2,3/2}$ when brought close to such a
metamaterial:
\begin{gather}
\ket{(nS_{1/2})^{1}} = C_{nS_{1/2}nS_{1/2}} \pc{z_A}
\ket{(nS_{1/2})^{0}} + \nonumber \\
\sum_{k} C_{nS_{1/2}kP_{1/2,3/2}} \pc{z_A} \ket{(kP_{1/2,3/2})^{0}}\,,
\end{gather}
In Fig.~\ref{Fig:Prob32S}, we show the probabilities for this
admixture as a function of atom-surface distance $z_A$.
It can be detected by probing an originally dipole-forbidden
transition $nS_{1/2}\to n' S_{1/2}$ which becomes weakly allowed due to 
surface-induced mixing:
\begin{multline}
\bra{(nS_{1/2})^1}
\hat{\mathbf{d}} \ket{(n'S_{1/2})^0} \\
= \sum_{k} C_{nS_{1/2}kP_{1/2,3/2}}^\ast \pc{z_A} \bra{(kP_{1/2,3/2})^0}
\hat{\mathbf{d}} \ket{(n'S_{1/2})^0}. 
\end{multline}
Its magnitude depends on the atom-surface distance and could
potentially be used as a measuring tool for determining $z_A$.

As an example of this effect, take the state $32S_{1/2}$ where, for
simplicity, we only consider the admixture with the state $32P_{1/2}$.
The transition dipole moment $\mathbf{d}_{32S,33S} (z_A)$ associated
with the now weakly allowed atomic transition $32S_{1/2} \to
33S_{1/2}$ is given by $C_1^* (z_A) \, \mathbf{d}_{32P,33S}$ and, from
Fig.~\ref{Fig:Prob32S}, one can see that the ratio $d_{32S,33S} (z_A)
/ d_{32S,32P}$ can be as large as $15 \%$.

\section{Summary}
\label{sec:finalremarks}

Within the framework of macroscopic QED, we have shown that the
failure of perturbation theory for highly excited Rydberg atoms
interacting with a nearby surface can be overcome by calculating their
energy shifts from an exact diagonalization of the interaction
Hamiltonian. In the case of very small atom-surface distance or very
large principal quantum number $n$, the deviation of the exact energy
shift from the second-order perturbation theory result can be
appreciable. We have further shown that, contrary to expectations
from single-mode coupling, surface-induced state mixing is suppressed
for good conductors despite large interaction energies.  We have shown
that in the particular example of a metamaterial with low-frequency
resonances, an atom acquires a finite probability to be found in a new
internal energy eigenstate when brought close to a surface.

\acknowledgements
SR is supported by grant SFRH/BD/62377/2009 from FCT, co-financed
by FSE, POPH/QREN and EU. SYB was supported by the DFG (grant BU
1803/3-1) and the Freiburg Institute for Advanced Studies. SS
acknowledges support by the DFG (grant SCHE 612/2-1).


\end{document}